\documentclass[twocolumn,nofootinbib]{revtex4-1}
\usepackage{latexsym,epsfig,amssymb, amsmath,nicefrac}

\usepackage{epsfig}

\usepackage{amsfonts,amsthm} 
\usepackage{bm,bbm}
\usepackage{graphicx}
\usepackage{esint}
\usepackage{color}
\usepackage{braket}
\def\K{K{\"a}hler}

\newcommand{\be}{\begin{equation}}
\newcommand{\ee}{\end{equation}}
\newcommand{\bea}{\begin{eqnarray}}
\newcommand{\eea}{\end{eqnarray}}
\def\bs{\begin{subequations}}
\def\es{\end{subequations}}

\newcommand{\rf}[1]{(\ref{#1})}

\begin{document}

\title{{\Large Natural Inflation in Supergravity and Beyond}}

\author{\bf  Renata Kallosh,  Andrei Linde and Bert Vercnocke}

\affiliation{Department of Physics and SITP, Stanford University, \\ 
Stanford, California 94305 USA\\ kallosh@stanford.edu, alinde@stanford.edu, bert.vercnocke@uva.nl}

\begin{abstract}
Supergravity models of natural inflation and its generalizations are presented. These models are special examples of the class of supergravity models proposed in \cite{Kallosh:2010ug}, which have a shift symmetric \K\, potential, superpotential linear in goldstino, and  stable Minkowski vacua.  We present a class of supergravity models with arbitrary potentials modulated by sinusoidal oscillations, similar to the potentials associated with axion monodromy models. We show that one can implement natural inflation in supergravity even in the models of a single axion field with axion parameters $O(1)$. We also discuss the irrational axion landscape in supergravity, which describes a potential with infinite number of stable Minkowski and metastable dS minima.

\end{abstract}

\maketitle

\section{ Introduction}

It appears that one of the popular models of inflation, called natural inflation \cite{Freese:1990rb}, which was proposed 24 years ago, has not yet been generalized to supergravity with stabilization of all moduli. The goal is to find a supergravity model that would lead to the natural inflation potential of the axion field $\phi$
\be
V=\Lambda^4 \Bigl(1-\cos{\phi\over f}\Bigr)
\label{Natural}\ee
with Minkowski minimum at $\phi=0$.
The supergravity axion valley models proposed and studied in \cite{Kallosh:2007ig,Kallosh:2007cc}, and used more recently in \cite{Czerny:2014qqa}, almost did the job. They have the following \K\, potential and superpotential
 \be\label{ax}
 K= {(T+\bar T)^2\over 2}\, ,  \qquad W= W_0 + A e^{-aT} + B e^{-bT} \ .
\ee
The real part of the modulus is stabilized in this model and the imaginary part plays the role of the light axion $\phi$. The resulting potential is almost of the form  \rf{Natural}. However, in this class of models
the minimum of the potential is in AdS space. Therefore one has to specify an uplifting procedure, which uplifts the AdS minimum to a Minkowski one, or even better, to a de Sitter minimum with a tiny cosmological constant. Various uplifting procedures have been proposed over the years, but some of them cannot be described at the supergravity level, whereas some others may lead to modification of the functional form of the potential upon uplifting. 
As a result, to the best of our knowledge, explicit supergravity models realizing such an uplifting in a way consistent with moduli stabilization and leading to natural  inflation \rf{Natural} are still unavailable. For a recent discussion of the axion inflation models see for instance \cite{Baumann:2014nda} and  \cite{Pajer:2013fsa}.

The  purpose of this note is to present a very simple supergravity model with non-negative potential which upon stabilization of the non-inflaton moduli produces the natural inflation potential \rf{Natural}. It will be achieved in the context of the general class of models \cite{Kallosh:2010ug} describing chaotic inflation in supergravity. This class of models generalized the supergravity realization of the simplest chaotic inflation scenario ${m^{2}\over 2}\phi^{2}$ proposed in \cite{Kawasaki:2000yn}. 

The class of models developed in \cite{Kallosh:2010ug} has a built-in  feature which makes the potential non-negative.  The superpotential in these models is linear in the goldstino superfield $S$, whereas the \K\, potential is some function of either $\Phi + \bar \Phi$ or $\Phi + \bar \Phi$,
 and of $S\bar S$: 
\be
W = S f(\Phi) \ , \qquad K = K ((\Phi\pm \bar \Phi)^2, S\bar S) \ .
\ee
The \K\, potential $K ((\Phi\pm \bar \Phi)^2, S\bar S)$ does not depend on one of the combinations $(\Phi\mp \bar \Phi)$, which plays the role of the inflaton field in this scenario. If one can stabilize the field $S$ at $S = 0$, then $W = 0$, and the potential becomes manifestly non-negative: 
\be
V= e^K (|DW|^2 -3W^2)|_{S=0}= e^K \partial_S W  \partial_{\bar S} \bar  W \geq 0 \ .
\ee
If, in addition, one can ensure that one of the combinations of the fields $(\Phi\pm \bar \Phi)$, which is orthogonal to the inflaton field, vanishes during inflation, then the inflaton potential becomes 
\be
V= |f(\Phi)|^2 \ .
\label{general}\ee
The required stabilization conditions are rather mild, which allows to have a functional freedom in the choice of the inflaton potential in supergravity  \cite{Kallosh:2010ug}.

As we will see, this class of models can easily incorporate natural inflation. Moreover, by a simple extension of the supergravity versions of natural inflation, one can find a family of positive definite inflationary potentials of arbitrary shape modulated by sinusoidal oscillations. These potentials are similar to the string theory inflaton potentials associated with axion monodromy \cite{Silverstein:2008sg,Flauger:2009ab}.

\section{Natural inflation in supergravity}

We  discuss various supergravity embeddings of natural inflation and related models. They all depend on two complex fields: $T$ and a goldstino $S$. Following the discussion above, we will use K\"ahler potentials which depend on either of the combinations $T\pm \bar T$, of the form:
\begin{equation}
 K^\pm = \pm{(T\pm\bar T)^2\over 2} + S\bar S - g (S\bar S)^2\ .
\end{equation}
The term $g (S\bar S)^2$ is introduced for stabilization of the field $S$ at $S = 0$, and the inflaton is the combination $T\mp \bar T$ not appearing in the K\"ahler potential.\\

{\bf  Model 1}

The superpotential and K\"ahler potential are
\be
W=\frac {\Lambda^2} {\sqrt 2} S (1- e^{-aT}) \, ,\quad K= K^+\ .
\label{model1}\ee
For convenience we introduce the canonically normalized real fields $\phi,\beta,s,\alpha$:
\begin{equation}
 T = \frac{\beta + i \phi}{\sqrt 2}\,, \quad S = \frac{\alpha + i s}{\sqrt 2} \ .
\label{compl}\end{equation}
We find that the potential has a minimum   at $S=0$ and  $T+\bar T=0$ and at $a\phi/\sqrt 2 = 2 \pi n$.

We have computed the masses of the fields $\beta,\alpha,s$ and have found that the stability analysis of [1] applies: the masses of the fields are of the order of the Hubble parameter during slow roll inflation, under the condition that $g\gtrsim 1/12$. Namely,
\bea
\frac{m^2_{\beta}}{H^2}&=& 6+\frac 32 a^2 + \frac 34 a^2\left(\sin\frac{a \phi}{2\sqrt{2}}\right)^{-2},\\
\frac{m^2_{s}}{H^2}&=&12g + \frac34 a^2\left(\sin\frac{a \phi}{2\sqrt{2}}\right)^{-2}.
\eea
Since the potential only depends on $S \bar S = s^2 + \alpha^2$, $\alpha$ and $s$ have the same mass.

Thus, during inflation the field $T+\bar T$ is heavy and quickly reaches its minimum at $T+\bar T=0$. The field $S$ is also heavy, for  $g\gtrsim 1/12$, and also vanishes. However, one may have an interesting scenario even if one discards the stabilization term  $g (S\bar S)^2$. Then the field $S$ remains light, and its perturbations can be generated during inflation. If the field $S$ rapidly decays at the end of inflation, these fluctuations remain inconsequential. However, if it is stable, or decays long after the end of inflation, one can obtain isocurvature fluctuations, or additional adiabatic perturbations via the curvaton mechanism \cite{Demozzi:2010aj}. The inflaton field $ \phi$ remains light and has the following potential 
\be
V|_{S=0, \, T+\bar T=0}=  \Lambda^4 \Big (1-\cos { a\phi \over \sqrt 2} \Big) \ ,
\ee
in agreement with \rf{general}.

We present the picture of the potential during inflation in Fig.\ 1.
\begin{figure}[h!t!]
\begin{center}
\hskip -0.25cm  \includegraphics[scale=0.31]{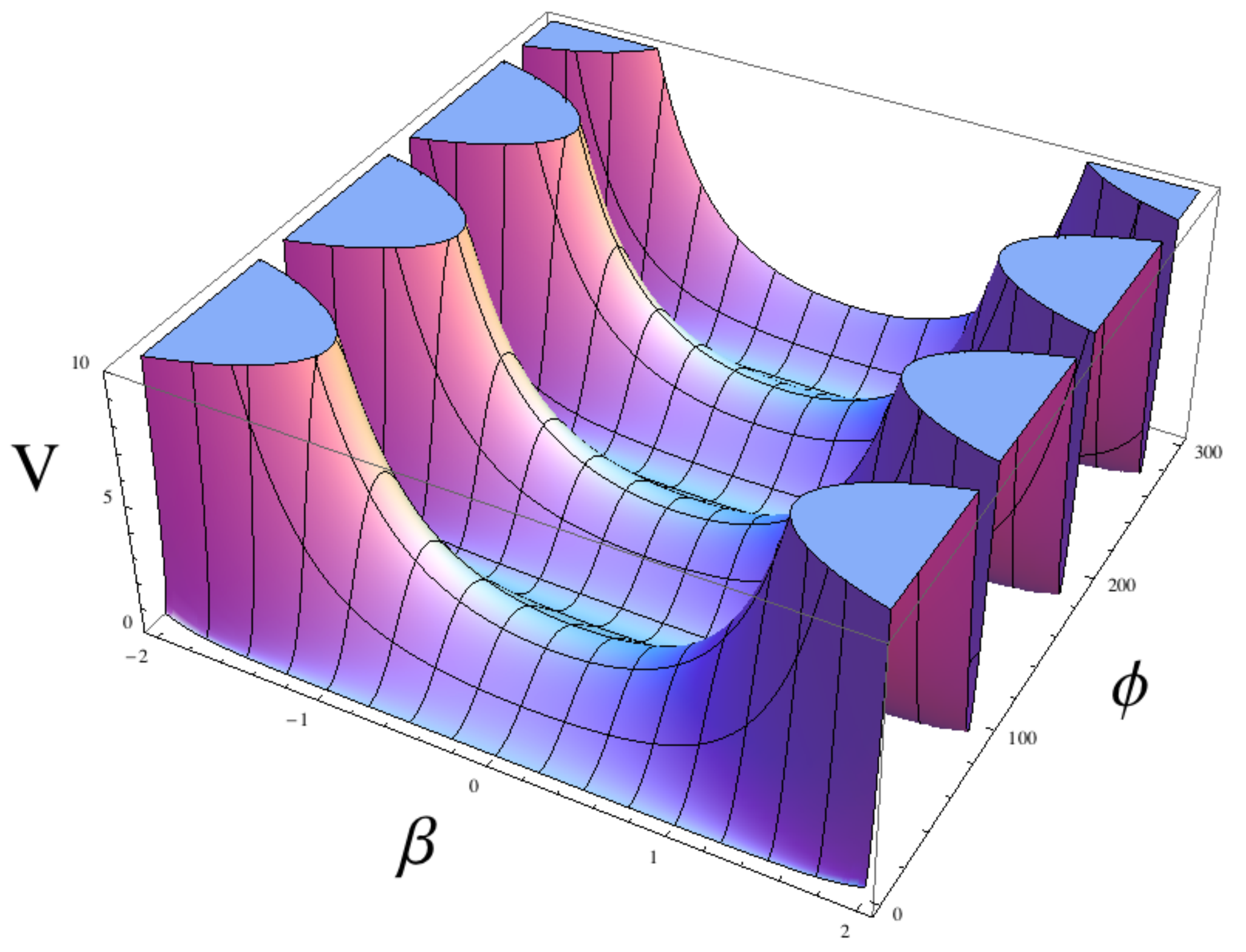}
\end{center}
\caption{\footnotesize Potential of the natural inflation model \rf{model1} in supergravity at $S=0$ and $T=\frac{\beta + i \phi}{\sqrt 2}$.  During inflation $\beta=0$ at its minimum and $\phi$ is the inflaton field with the potential \rf{Natural}. This plot is made for  $a=0.1$. All fields are given in Planck units, and the potential is in units  $\Lambda^4$.}
\label{chi0b}
\end{figure}

\

{\bf Model 2}

The superpotential and K\"ahler potential are
\be
W=\sqrt 2 \Lambda^2 S \sin \frac{aT}2  \, ,\quad K= K^-  \ .
\label{model2}\ee

During inflation the bosonic stabilized model is the same as Model 1 for the canonically normalized real fields
\begin{equation}
 T = \frac{\phi + i \beta}{\sqrt 2}\,, \quad S = \frac{s + i \alpha}{\sqrt 2}\ .
\end{equation}
with the inflaton potential 
\be
V|_{S=0, \, T+\bar T=0}=2 \Lambda^4  \Big (\sin{ a\phi \over {2\sqrt 2 }}\Big )^2= \Lambda^4  \Big (1- \cos { a \phi \over \sqrt 2}\Big ) \ .
\ee

However, in general, Model 2 is slightly different, and there is a small difference  in masses of the stabilized fields:
\bea
\frac{m^2_{\beta}}{H^2}&=& 6+ \frac 34 a^2\left(\sin\frac{a \phi}{2\sqrt{2}}\right)^{-2},\\
\frac{m^2_{s}}{H^2}&=&12g -\frac 34 a^2+ \frac34 a^2\left(\sin\frac{a \phi}{2\sqrt{2}}\right)^{-2}.
\eea
The potential is very similar to the one of  model 1 shown in Fig.\ 1.

\

{\bf Model 3}

Here we show how the replacement of all scalars of the type $z\rightarrow iz$ works when we start with Model 2 and create this Model 3. We start with Model 2 in \rf{model2} in the form
\be
W= \frac{ i \Lambda^2}{  \sqrt 2} S ( e^{-i aT} - e^{ i aT} \Big ),\, \quad K= K^- 
\label{model2'}\ee
and perform the following change of variables $T  \rightarrow i T$ and $S  \rightarrow   i S$. We find
\be
W= {\Lambda^2 \over \sqrt 2}S ( e^{-aT} - e^{ aT} \Big ) ,\quad K= K^+   \ ,
\label{model2a}\ee
where the inflaton is now the imaginary part of a scalar $T$.
It leads to exactly the same physics as Model 2, and very similar physics compared to Model 1. The relevant potential is, therefore, given again (approximately) by Fig.\ 1.

\

{\bf Model 4 }

Finally we give a supergravity model reminiscent of a potential with a sum of several cosines as in \cite{Kallosh:2007cc,Czerny:2014qqa}.
The superpotential and K\"ahler potential are
\be
W=\sqrt 2 \Lambda^2 S \left(A\sin \frac{aT}2+ B\sin \frac{bT}2\right), ~~~~
K= K^- \ .
\label{model2aa}
\ee
The inflaton potential is 
\be
V=2 \Lambda^4  \Big (A\sin{ a\phi \over {2\sqrt 2 }}+B\sin{b\phi \over {2\sqrt 2 }}\Big )^2 \ .
\ee

\section{Irrational axion landscape}

Now we will make what could seem a minor modification of the previous model, but we will find a dramatically different potential. The \K\ potential now is $K= K^{+}$, and the superpotential slightly differs from \rf{model2a} 
\be 
W={\Lambda^2}  S (1- Ae^{-aT}- Be^{-bT}) \ .
\ee
The potential at $S=T+\bar T= 0$ is
 \bea
&V&= {\Lambda^4} \Bigl(1+A^2 +B^2 - 2A\cos { a \phi \over \sqrt 2} \nonumber\\
&+&2AB \cos { (a-b) \phi \over \sqrt 2} -2B \cos { b \phi \over \sqrt 2}\Bigr)\ .
 \eea
\begin{figure}[h!t!]
\begin{center}
 \includegraphics[scale=0.65]{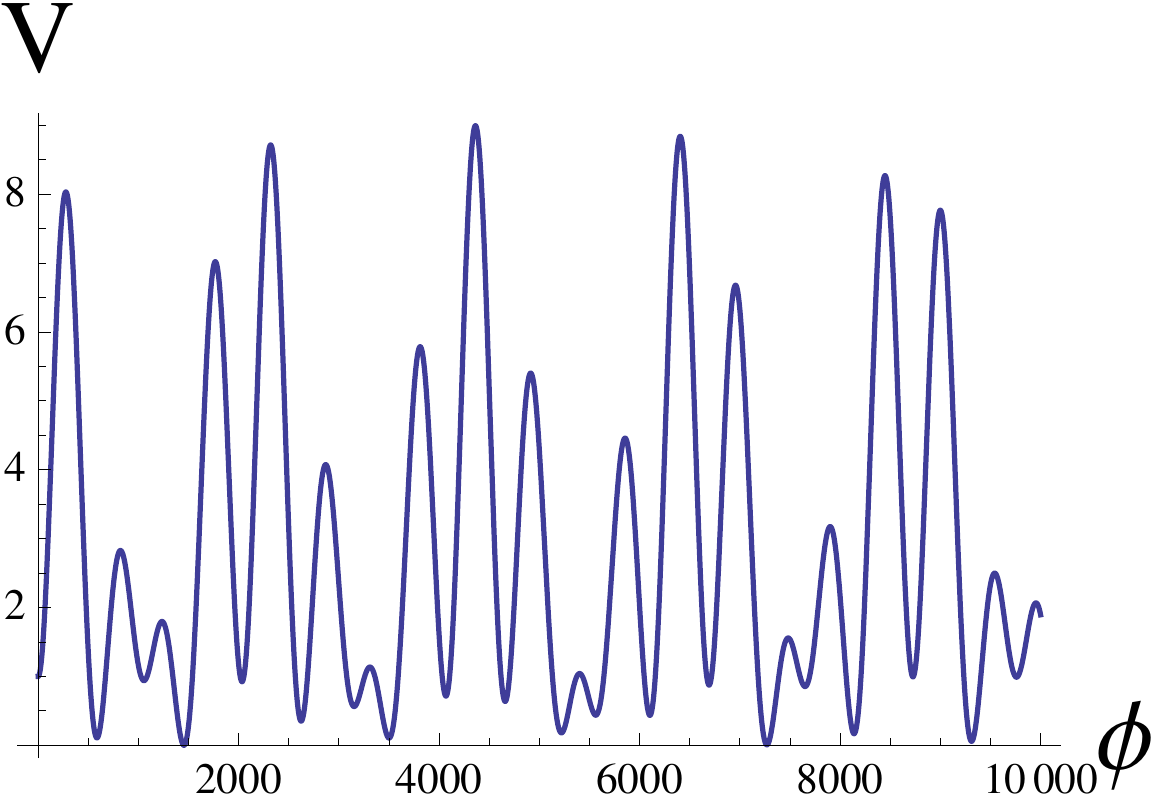}
\end{center}
\caption{\footnotesize Irrational axion potential for $A=B = 1$, $a = 0.01 \sqrt3$, $b = 0.005\sqrt 7$. The field is shown in Planck units, from 0 to 10000. This may create an impression that the potential is very steep, but in fact the potential is very flat and allows chaotic inflation. Just as in the string landscape scenario \cite{Linde:1986fd,Lerche:1986cx,Bousso:2000xa,Kachru:2003aw,Douglas:2003um,Susskind:2003kw}, inflation may end in any of the infinitely many metastable dS vacua with different values of the cosmological constant  \cite{Banks:1991mb}.}
\label{irr}
\end{figure}

This potential has an interesting behavior discussed in  \cite{Kallosh:2007cc,Czerny:2014qqa}, but now we have its explicit supergravity implementation without any need for an uplifting. As discussed long ago by Banks, Dine and Seiberg \cite{Banks:1991mb}, a particularly rich behavior is possible if the ratio ${a\over b}=q$ is irrational. This leads to a landscape-type structure of the potential with infinite number of different stable Minkowski vacua and metastable dS vacua with different values of the cosmological constant, see Fig.\ \ref{irr}. If one of the constants $a$ and $b$ in this scenario is irrational, we have an infinite number of possible dS minima, which allows to solve the cosmological constant problem using anthropic considerations. 

Moreover, inflationary predictions in this scenario depend on the behavior of the inflaton potential in the vicinity of each of these dS vacua.
As a result, one can have a broad spectrum of possibilities which allows to fit a large variety of observational data within the context of a single model with a small number of parameters.

\section{Inflation for $a, b \gtrsim 1$}

Until now, we discussed the scenario with $a, b\ll 1$. However, string theory suggests that the parameters $a,b \gtrsim 1$. Can we still have natural inflation in that case?

Let us consider a model with 
\be
W={\Lambda^2}  S (e^{-aT}- e^{-bT}), ~~~~ K= K^{+}  \ .
\label{onefieldlarge}
\ee
We will assume that $a,b \gtrsim 1$, $a-b \ll 1$. One can show that in this case inflation is indeed possible. 

The absolute minimum of the potential in this theory is at $T = 0$. However, one can show that inflation occurs in the regime of a slow roll from the saddle point of the potential with ${\rm Re}\ T = a/2$.  ${\rm Re}\ T$ remains very close to $a/2$ during inflation, and only in the very end it starts moving towards the global minimum with $T =0$. The inflaton potential in this theory is well approximated by
\be
V=2 \Lambda^4 e^{-a^{2}/2} \, \Big (1-\cos {(a-b)\, \phi\over \sqrt 2}\Big ) \ .
\ee 
This potential allows inflation even for $a,b\gg1$ if the difference between $a$ and $b$ is small, $|a-b| \ll 1$. A similar idea, in a different context, was used in the racetrack inflation model \cite{BlancoPillado:2004ns}, and then applied to natural inflation in \cite{Kallosh:2007cc}. In this way, one can bring natural inflation one step closer towards its implementation in string theory. Note that we were able to do it in the theory of a single axion field.

\begin{figure}[h!t!]
\begin{center}
\hskip -0.4cm \includegraphics[scale=0.3]{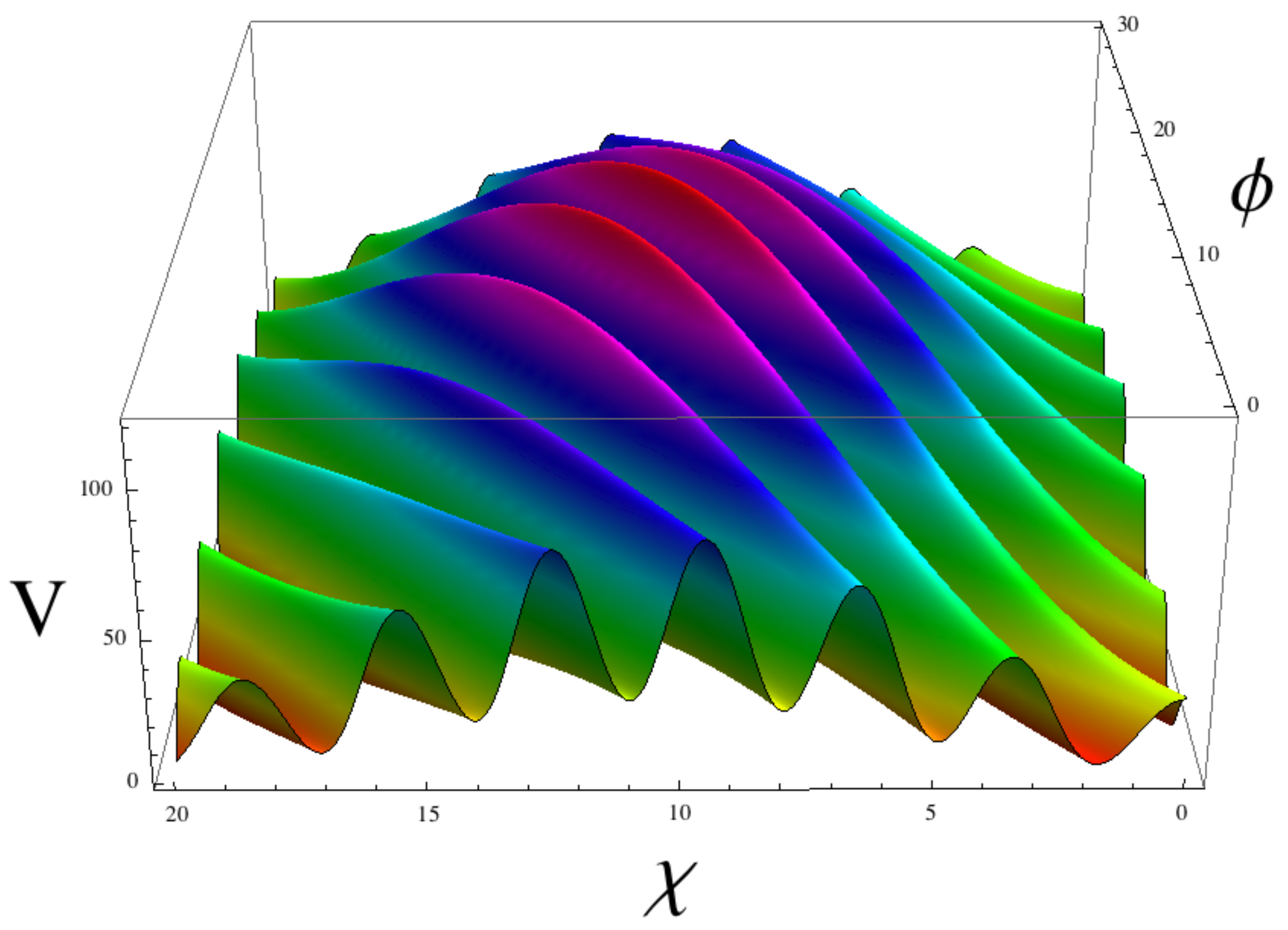}
\end{center}
\caption{\footnotesize Inflationary potential for natural inflation in the theory of two scalar fields, $T=\frac{\beta + i \phi}{\sqrt 2}$ and $U=\frac{\gamma + i \chi}{\sqrt 2}$, as a function of the fields $\phi$ and $\chi$. The potential shown in units of $\Lambda^{4}$, with all other parameters of the superpotential $O(1)$. Nevertheless, flat inflationary  valleys are formed for $|a-b| \ll 1$ or $|c-d|\ll1$. }
\label{chi0ca}
\end{figure}

A similar mechanism may work in the inflationary theory of many axion fields \cite{Kim:2004rp}. Until now, the multi-axion natural inflation scenario has not been implemented in supergravity. The simplest way to do it is to consider two axion fields, $T=\frac{\beta + i \phi}{\sqrt 2}$ and $U=\frac{\gamma + i \chi}{\sqrt 2}$, with the superpotential 
\be 
W={\Lambda^2}  S (Ae^{-aT}+ Be^{-bT}+C^{-cU}+ De^{-dU})  
\ee 
and the \K\ potential
\be
K = {(T+\bar T)^2\over 2} + {(U+\bar U)^2\over 2} + S\bar S - g (S\bar S)^2 \ .
\ee
For a proper choice of parameters, the potential of the fields has inflationary flat directions, as shown in Fig. \ref{chi0ca}.

Thus one can have inflation in such models as well. However, a full description of inflation in multi-axion models in supergravity can be rather involved. In general, all fields, including their real and imaginary parts, may evolve simultaneously during inflation, which makes investigation of inflation in such models more complicated than in the simple single-inflaton field \rf{onefieldlarge}.

\section {Modulated chaotic inflation potentials}

Here we propose supergravity models closely related to  the explicitly bosonic models in \cite{Flauger:2009ab} for oscillations in the CMB from axion monodromy inflation.
We take a generic function $f(T)$ in the superpotential complemented by some sinusoidal modulation of the form
\be
W= S\Bigl[f(T) + A\sin (aT)\Bigr],\quad K= K^-\ .
\ee
Here $T=\frac{\phi + i\beta}{\sqrt 2}$. If needed, one can also add to the \K\, potential the stabilization term $(T+\bar T)^2  S\bar S$ for stabilization of $\beta$, but usually it is not required  \cite{Kallosh:2010ug}. For $S = \beta = 0$, one finds the  inflaton potential
\be
V = \Bigl[f\Bigl(\frac{\phi}{\sqrt 2}\Bigr) + A\sin \frac{a\phi}{\sqrt 2}\Bigr]^2 \ .
\ee

\begin{figure}[h!t!]
\begin{center}
 \includegraphics[scale=0.35]{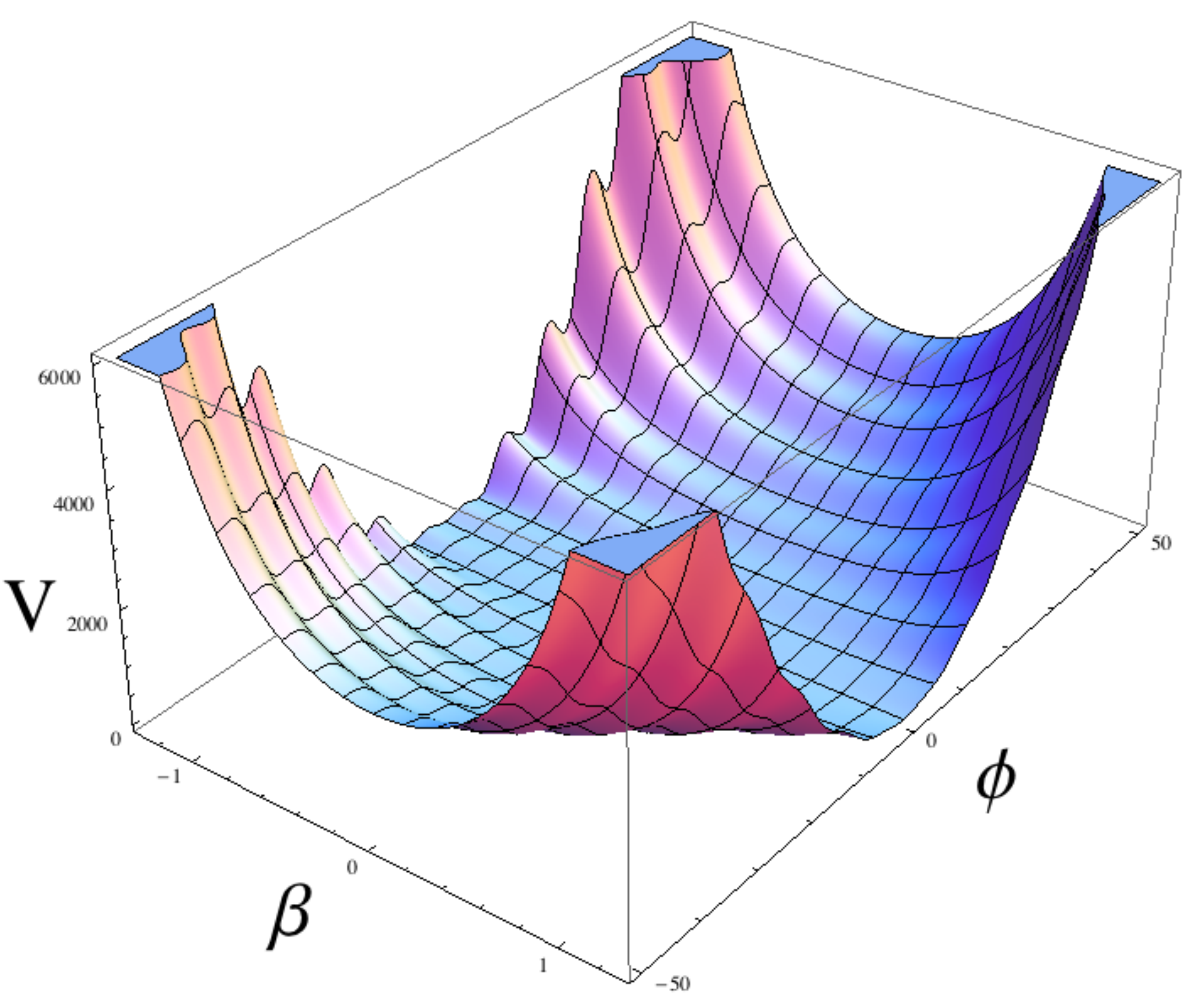}
\end{center}
\caption{\footnotesize Potential of the modulated chaotic inflation in supergravity \rf{model1a} at $S=0$ and $T=\frac{\beta + i \phi}{\sqrt 2}$, for $a = 1$, $b = 1.3$. It is similar to the potentials encountered in axion monodromy models \cite{Silverstein:2008sg,Flauger:2009ab}. Fields are shown in Planck mass units, the scale of the potential is in units $\Lambda^{4}$.}
\label{chi0c}
\end{figure}
In the limit when $f\Bigl(\frac{\phi}{\sqrt 2}\Bigr) \gg A\sin \frac{a\phi}{\sqrt 2}$ the modulation of the inflaton potential is small and  we find
\be
V \approx f^{2}\Bigl(\frac{\phi}{\sqrt 2}\Bigr)+ 2 A\, f\Bigl(\frac{\phi}{\sqrt 2}\Bigr) \sin \frac{a\phi}{\sqrt 2} \ .
\ee
It is only slightly different from potentials with modulation studied in the literature, see  \cite{Flauger:2009ab} and references therein.
They  assumed that the amplitude of modulation is constant, whereas in our case it is proportional to $f\Bigl(\frac{\phi}{\sqrt 2}\Bigr)$. The difference is not crucial because $f\Bigl(\frac{\phi}{\sqrt 2}\Bigr)$ may not change much on scales studied by the CMB observations.

A similar scenario can be also implemented in a different context. We can consider, for example, the following supergravity model which produces a quadratic axion potential with sinusoidal modulations:
\be
W={\Lambda^2}  S (1- e^{-aT}+ bT) ,\quad K= K^+\ . 
\label{model1a}\ee
We plot the potential $V$ at $S=0$ in Fig.\ \ref{chi0c}.
The potential as a function of the inflaton is
\be
V = {\Lambda^4}\left({b^2\over 2} \phi^2 +  \sqrt 2 \, b \phi \, \sin {a\phi\over \sqrt 2} + 4 \sin^2 {a\phi\over \sqrt 2} \right)\ .
\label{modulations}\ee

\

In conclusion, we have presented here a supersymmetric version  of natural inflation \cite{Freese:1990rb} and of the models with arbitrary potentials modulated by sinusoidal oscillations, similar to the potentials associated with axion monodromy models \cite{Silverstein:2008sg,Flauger:2009ab}. The corresponding supergravity models are simple and have Minkowski vacua. We have shown that one can implement natural inflation in supergravity even in the models of a single axion field with axion parameters $O(1)$. Embedding of the irrational axion models \cite{Banks:1991mb} in supergravity allows many stable Minkowski vacua and metastable dS vacua with different values of the cosmological constant. It would be interesting to explore a possible relation of such supergravity models to the string theory landscape.

We are grateful to Diederik Roest, Eva Silverstein, Alexander Westphal and Timm Wrase for discussions of closely related issues.  This  work is supported by the NSF Grant PHY-1316699 and SITP.  The work of RK is also supported by the Templeton grant ``Quantum Gravity frontier''.  BV gratefully acknowledges the Fulbright Commission Belgium for financial support.

\providecommand{\href}[2]{#2}\begingroup\raggedright\endgroup

\end{document}